\documentstyle[psfig]{caosp}            
\begin{document}
\pubyear{1993}
\volume{23}
\firstpage{1}
\title{Some recent results for the roAp stars}
\author{D. W. Kurtz \inst{}}
\institute{Department of Astronomy, University of Cape Town, Rondebosch 7700,
South Africa}

\date{December 2, 1997}
\maketitle
\begin{abstract}
Three topics are discussed: 1) Photometric observations of the rapidly
oscillating Ap stars have shown that the pulsation amplitude drops
dramatically as a function of wavelength from the blue to the red.
A theoretical derivation, plus modelling, indicate
s that this is because the vertical wavelength of the pulsation mode is short
compared to the scale height of the atmosphere; in fact, it indicates that we
are seeing a pulsation node in the observable atmosphere. Radial velocity
observations, and theoretical calculations now support this. The implication
for other research on CP stars is that this can provide observational
constraints on the atmospheric structure independent of traditional spectral
analysis. 2) Luminosities of roAp stars can be determine
d from asteroseismology. A recent comparison of such asteroseismic
luminosities with HIPPARCOS luminosities is shown. This suggests that roAp
stars have lower temperatures and/or smaller radii than previous models have
used, or that the magnetic fields in these stars alter the frequency
separations. 3) The latest results of our long-term monitoring of the
pulsation frequencies in certain roAp stars are discussed. There is a clear
cyclic variability to the pulsation cavity, hence the sound speed and/or 
sound travel time (radius) of these stars. This might be indicative of
magnetic cycles at a level that magnetic measurements cannot currently
detect, although there is no theoretical support for such an idea.

\end{abstract}
\section{Introduction: Of snakes and fish}
High in the mountains of Angola arises the Kavango river. It flows down
across the deep sands of the Kalahari desert in Botswana where it spreads out
over 200 km to form the Okavango Swamps - one of the greatest wildlife
refuges in Africa. In the Swamps the water meanders through papyrus-choked
channels dotted with small desert islands. The sand filters the water to
spectacular clarity and purity. Hippos wallow in the main channels;
crocodiles are common. On the biggest island, Chief's Island, the
``real'' Africa of the imagination comes alive: There are elephants, lions,
cape buffalo, impala, warthogs, and the ``Swamp Specials'', Tssessebe and
Lechwe - buck specially adapted to swamp living.

Many years ago five friends and I from Cape Town flew deep into the Swamps
where we joined three Batswana guides in dugout canoes (called Mekoros) for
a 10-day camping trip through the Swamps. We had no one common language
between our three guides and the six of us, but amongst us all we spoke enough
of a mixture of Tsetswana, Zulu, Xhosa, Fanikalo, English and Afrikaans to
communicate. Each night, around the campfire over a shared dinner, we told
stories of Africa. 

On our first night we six from Cape Town all crawled into warm, goose-down
sleeping bags - the latest and best in Western camping gear - snug against
the sub-freezing cold of a winter's cold snap. We lay on our groundsheets
oriented radially away from the fire with our feet at a safe distance from
the sparks which might damage the expensive sleeping bags, and our heads out
in the darkness where we could see the spectacular African Sky through the
foliage of the bushveld trees. Our three guides had only two blankets for the
three of them. One they put down in the sand to sleep on; they then curled up
together for warmth and put the other blanket over them. Their heads were
almost in the fire, and a stack of wood for stoking the fire during the night
was nearby. 

Even with the cold we asked them, ``Why do you sleep with your heads so close
to the fire? Aren't you afraid of being burned by the sparks?'' And they
explained, ``We have lived here all our lives. When a hyena or lion comes out
of the night and tries to bites us, we would rather it bit at our feet, than
our heads. We will then sit up and hit it with this Panga [machete] to scare
it off.'' ``Ha, ha, ha'', we laughed, ``Listen to our guides trying to scare
us city-slickers; well, you can't scare us! We are experienced campers.''

As we were trying to go to sleep there came the nearby roar and clatter of a
train approaching! How could this be? There are no trains in the Okavango
Swamps. Our guides patiently explained that it was a herd of Lechwe fleeing at
high speed through the shallows of the swamp, possibly being chased by lions.
We fell asleep thinking ``there they go - trying to scare us again. Ha Ha.'' 

But then, in the night, the lion roars came. They sounded like they were just
beyond the shadows of the firelight - very near to our exposed heads. When
you hear lions roar nearby, there is a louder-than-possible, low-frequency,
or even sub-sonic, rumble which shakes your insides and turns them to jelly.
It says, ``Be Frightened!'' And you are frightened. When we awoke the next
morning all six Capetonians were sleeping with feet away from the fire and
heads nearly in it; and that is how we slept the rest of the trip.

One night around the campfire we were exchanging our multiglottal stories.
Our guides told us of a giant snake which lives far up near the headwaters
of the Kavango River in Angola, a snake so big that it can swallow a Mekoro
and its three occupants whole in a single swallow! When I recovered from
laughing uproariously at this ridiculous claim, I asked ``Have $you$ ever
seen one of these snakes?'' ``Wellllll, actuallllllly, hmmmm, no we
haven't! .... BUT! We have reliable friends who have seen them, and
we $know$ they are there.'' 

I just poo-pooed the idea. I said it was absurd. There is no snake in the
world anywhere near that big.

Then I decided to tell the guides about the Great White Sharks of False Bay
near Cape Town. In the Okavango Swamps lives the terrifying and terrific
Tigerfish, with its razor teeth and mouth big enough to take off the hand of
an unwary or foolish fisherman. Our guides knew this fish well; they all knew
about the huge sea, too, although they hadn't seen it. False Bay is famous
for its 5-m Great Whites. So I told our guides that in the sea near Cape Town
we have fish so big that it can swallow a person in 
only two bites!

``Hahahahahahahahahahahahahahahahahahahahahahahahaha!!!!!!!!!!!!''

All three guides were rolling on the ground in the sand with tears streaming
down their faces. When they finally recovered from laughing uproariously at
this ridiculous claim, one looked at me, trying to control involuntary
outbreaks of more laughing, and asked ``Have $you$ ever seen one of these
fish?'' 

I was rather taken aback by the question. I thought about it, then said in all
honesty, ``Wellllll, actuallllllly, hmmmm, no I haven't! .... BUT! I have
reliable books! I have seen reliable films! I have reliable friends who have
seen them, so I $know$ they are there.''

They just poo-pooed the idea. They said it was absurd; there is no fish in the
world anywhere near that big.

Both ``the fish'' and ``the snake'' were undoubtedly-real and obviously-absurd,
depending on our preconceptions and viewpoints. In science a major task is to
discriminate between the fish and the snake. Either may seem
plausible - indeed, may seem probable - and either may turn out to be absurd
in the end. Great care and hard work are needed to learn which is true, and
which absurd.

\section{Atmospheric nodes in rapidly oscillating Ap stars}

I was very excited by the announcement by Jaymie Matthews (Matthews $et~al.$
1990, 1996) that atmospheric T-$\tau$ in roAp stars could be determined from
the amplitude of their pulsations at different wavelengths. A problem much
discussed for the roAp stars is that of the critical frequency: Many of them
seem to pulsate with frequencies so high that their vertical wavelength should
be of order of the thickness of the surface boundary layer. In that case the
mode should become evanescent; without very large driving there is not enough
energy for modes with frequencies above the critical frequency to be
maintained as standing waves. This problem is discussed by Gautschy \& Saio
(1998) and Audard $et~al.$ (1998). Matthews found that the T-$\tau$ gradient in 
HR~3831 was much steeper than in a normal A star model; this steepness sharpens
the surface boundary and increases the critical frequency. It also results in
the derivation of smaller overabundances of the rare earths and lanthanides for
a given observed line strength.

Because this looked such a fruitful field, Thebe Medupe and I decided to
begin a large project to observe many roAp stars in $UBVRI$ colours to apply
Matthews' technique of inferring T-$\tau$ from limb darkening affects on the
pulsation amplitude. Instead of modelling this first on a computer, however,
we started with a simple analytic derivation of what we expected, making some
reasonable first-order assumptions, such as a black body energy distribution
and the Wien approximation. To our surprise we found a ``snake'': We found that
limb-darkening is not an important effect (Kurtz \& Medupe 1996, Medupe
\& Kurtz 1997). The pulsation amplitude drops from the blue to the infrared
by a factor of two greater than the drop expected for a simple black body.
Limb-darkening can only account for about 12.5\% of this. Much more
sophisticated computer modelling confirms this result. 

Medupe and I found that the observed amplitude is expected to be 

\begin{equation}
A_{\lambda \rm obs}=1.086\,\frac{1}{4}\sqrt{\frac{3}{\pi}}\,\left(
\frac{4-\mu_{\lambda}}{3-\mu_{\lambda}}\right)\cos\alpha \,
\frac{hc}{\lambda kT_0}
\frac{\Delta T}{T_0}
\end{equation}

\noindent where $\mu_{\lambda}$ is the wavelength-dependent limb-darkening
coefficient, $T_0$ is the temperature at the atmospheric level appropriate to
the wavelength observed, and $\Delta T$ is the semi-amplitude of the
pulsational polar temperature variation. What is clear is that only a strong
dependence on $\Delta T$ can match the observations. We suggested that we were
seeing over a large fraction of the vertical wavelength of a high-overtone
p-mode. 

Several studies now give strong support to this suggestion; it looks like it
is a ``fish''. Baldry $et~al.$ (1998) show radial velocity measures in which
an atmospheric node is evident in $\alpha$ Cir. Pulsation phase versus
equivalent width diagrams give further support in studies of $\alpha$ Cir
(Baldry $et~al.$ 1997) and $\gamma$ Equ (Kanaan \& Hatzes 1997). Theoretical
models are now also producing pulsation nodes in the observable atmospheres
of roAp stars. Fig. 1 shows computations kindly provided 
by Alfred Gautschy which show a temperature node of just the sort hypothesised
by Medupe and me in an A star model for a k=27, $\ell$=1 mode. We thus have
the prospect of a new asteroseismic technique for probing the atmospheres of
the most peculiar stars known. 

\begin{figure}[hbt]                                                           
\psfig{figure=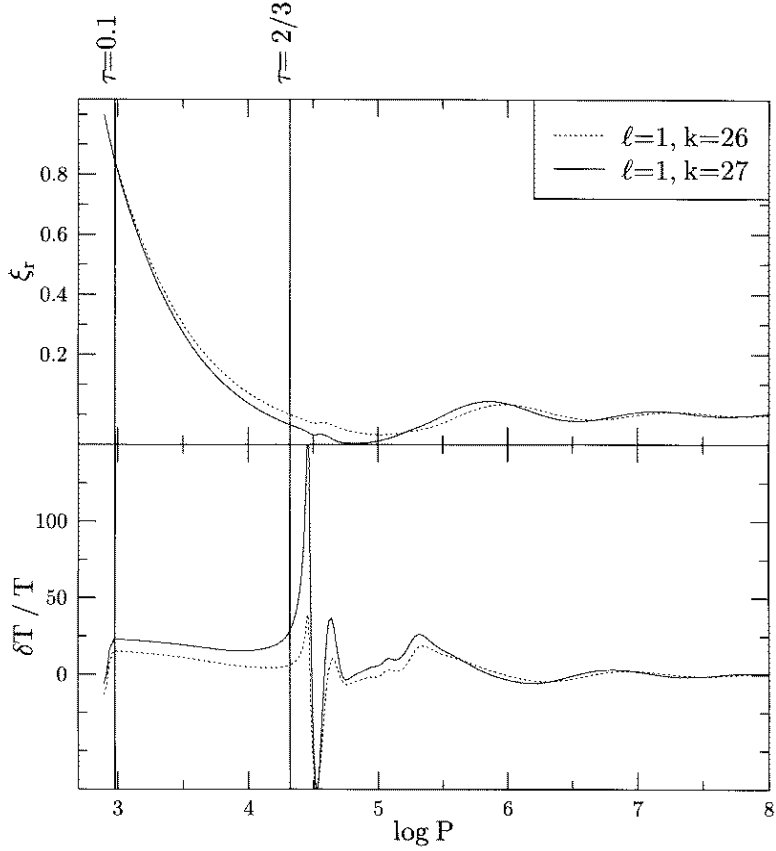,height=11.5cm}                                        
\caption{This diagram (courtesy of Alfred Gautschy) shows the sharp variation
of $\frac{\Delta T}{T_0}$ in the observed atmosphere for an A star model. This
is consistent with our explanation for the sharp drop in observed amplitude as
a function of wavelength.}
\end{figure}                                                                

\section{Asteroseismic Luminosities}

The frequency spacings of 12 roAp stars allow theoretical asteroseismic
estimates of their luminosities. These stars are spectroscopically so peculiar
that these luminosities have been thought to be the best available. However,
severe doubt was thrown on this by Dziembowski \& Goode (1996) when they
calculated that the magnetic perturbation to the pulsation frequencies is so
large (of order tens of $\mu$Hz) that perturbation theory will not work, and
the development of the oblique pulsator model ($e.g.$ 
see Shibahashi \& Takata 1993; Takata \& Shibahashi 1995) was on shaky ground.

\begin{figure}[htbp]                                                           
\centerline{\psfig{figure=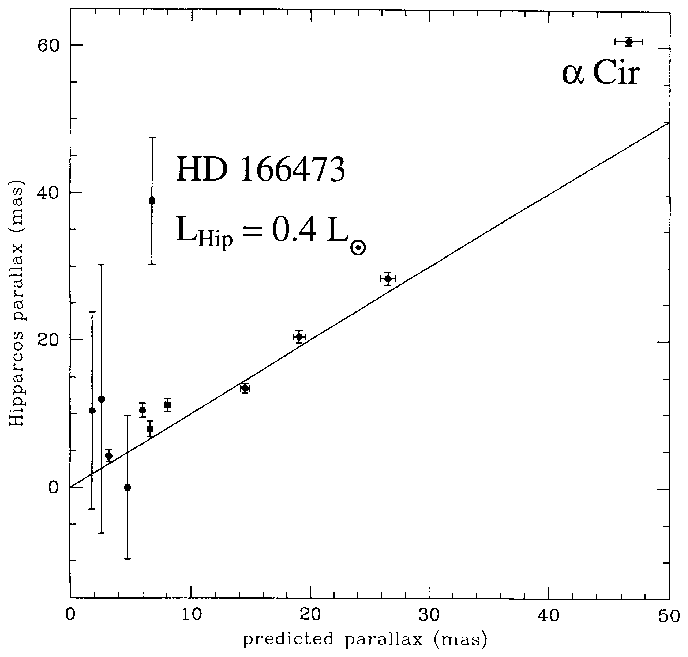,height=15cm}}
\caption{A comparison of the HIPPARCOS and asteroseismic parallaxes. The
asteroseismic parallaxes are systematically slightly smaller, and two stars
(discussed in the text) stand out. Otherwise, the agreement is remarkably
good - indicating that the asteroseismic luminosities are correct. }
\end{figure}                                                                

With the new HIPPARCOS parallaxes Matthews $et~al.$ (1997) have been able to
test the asteroseismic luminosities. Figure 2 is from their paper, and it
shows good agreement. There are two exceptions. One is HD 166473 which,
according to the HIPPARCOS parallax, has a luminosity only 40\% that of the
Sun. For a main sequence A star this is obviously incorrect, so we conclude
that the HIPPARCOS parallax is wrong. For $\alpha$ Cir the discrepancy is a
bigger problem. The asteroseismic parallax is determined from a frequency
spacing of 50 $\mu$Hz (Kurtz $et~al.$ 1994). The secondary frequencies in
this star give a strong indication that 50 $\mu$Hz is the right value, but
they are all of amplitude about 0.2 mmag, so this needs confirmation. In
this important case another intensive study of $\alpha$ Cir is called for.

For the remainder it can be seen from Figure 2 that the asteroseismic
parallaxes are systematically slightly smaller than the HIPPARCOS parallaxes.
That means that the A star models used to predict the luminosity from the
frequency separations systematically give too large a luminosity, hence the
models are too hot and/or too large in radius, or the magnetic field affects
the frequency $spacings$. To get an order of magnitude feel for the
discrepancy: If we assume the radii are correct, then the parallax 
disagreement indicates that the effective temperatures of the roAp stars are
about 1000 K cooler than the A star models from which the asteroseismic
luminosities were calculated. 

I remind you that the atmospheres of the roAp stars are peculiar to
pathological. Luminosities are notoriously difficult to determine, and even
the effective temperature can lead to decades-long, acrimonious dispute. In
the most extreme case, HD 101065, temperature estimates range from less than
6000 K to over 8000 K! This particular star is arguably the most peculiar in
the sky. In its visible spectrum the lead role is played by singly ionised
Holmium, with strong supporting roles from Dysprosium, Neodymium, Gadolinium,
Samarium, Lanthanum, $etc$. (presuming you know where ``$etc$.'' leads with
that series as a starter). Asteroseismology is providing unique constraints
on the structure of the Ap stars.

\section{Frequency variability in roAp stars}

A group from the University of Cape Town and the South African Astronomical
Observatory has been observing roAp stars on a long-term basis for frequency
variability for 6 years now. In this on-going project we get one hour of
observation of HR~3831 on each possible night over the approximately 8-month
season when it is observable. The reason for the emphasis on this star is that
it is the best-studied of the roAp stars, it has interesting rotational
amplitude and phase variability which we need to remove to study the frequency
variability, and the rotational variations are interesting subjects to study
in their own right. We also observe for one hour once per week two other roAp
stars, HD~134214 and HD~128898 ($\alpha$ Cir). HD~134214 is singly periodic 
with the shortest known pulsation period for the roAp stars, 5.65 minutes. The
very bright star $\alpha$ Cir is nearly circumpolar, and has a single large
amplitude pulsation mode with only small amplitudes for other modes and small
rotational sidelobes. The frequency variability of HR~3831 is discussed in
Kurtz $et~al$. (1997); we have not published O-C diagrams for the other two
stars recently. 

It is clear that there are variations in the pulsation cavities of these stars
on time-scales of years. For HR~3831 the variations can be characterised as
cyclic with a time-scale of 1.6 years. For $\alpha$ Cir the time-scale is
about 6 years, and HD~134214 is harder to characterise. The O-C diagrams can
be seen in Kurtz (1998). The variations in O-C cannot be easily explained as
Doppler shifts caused by companions; for HR~3831 many companions would need
to be hypothesised. In addition, the Ap stars have a very low incidence of
binarity, only about 20\% are in short period binary systems. So the frequency
variations are intrinsic, and they indicate a cyclic variability in the
acoustic cavity - this may be anything which affects the sound speed. One 
speculation we have made is that this indicates a magnetic cycle. The time
scale and amplitude of the frequency variations are similar to those which
are seen in the sun over the solar cycle. Magnetic fields in Ap stars are
thought to be fossil, however, rather than dynamo generated, so this
suggestion has not met with much approval; a ``snake'' is suspected here.
Whatever the physical mechanism that is at work, we have a new observational
phenomenon which will eventually tell us more of the inner workings of 
the roAp stars via asteroseismology.

\end{document}